\documentclass[12pt]{article}
\usepackage{graphicx}
\usepackage{amsmath}

\renewcommand{\Re}{\mathop{\rm Re\,}\nolimits}
\renewcommand{\Im}{\mathop{\rm Im\,}\nolimits}
\newcommand{\Arg}{\mathop{\rm Arg\,}\nolimits}

\title{Exact Solutions for Loewner Evolutions}

\author{Wouter Kager and Bernard Nienhuis\\
Institute for Theoretical Physics\\
University of Amsterdam\\
Valckenierstraat 65\\
1018 XE Amsterdam, the Netherlands\\
e-mail: kager@science.uva.nl and nienhuis@science.uva.nl \\
\\
and \\
\\
Leo P.\ Kadanoff\\
The James Franck Institute\\
The University of Chicago\\
5640 S.\ Ellis Avenue\\
Chicago IL 60637 USA\\
e-mail: leop@UChicago.edu}
\date{}

\begin{document}

\maketitle

\begin{abstract}

In this note, we solve the Loewner equation in the upper half-plane with
forcing function~$\xi(t)$, for the cases in which $\xi(t)$ has a power-law
dependence on time with powers $0$, $1/2$ and~$1$. In the first case the
trace of singularities is a line perpendicular to the real axis. In the
second case the trace of singularities can do three things. If $\xi(t)=
2\sqrt{\kappa t}$, the trace is a straight line set at an angle to the real
axis. If $\xi(t)=2\sqrt{\kappa(1-t)}$, as pointed out by Marshall and
Rohde~\cite{MR}, the behavior of the trace as~$t$ approaches~$1$ depends on
the coefficient~$\kappa$. Our calculations give an explicit solution in which
for $\kappa<4$ the trace spirals into a point in the upper half-plane, while
for $\kappa>4$ it intersects the real axis. We also show that for $\kappa=9/2$
the trace becomes a half-circle. The third case with forcing~$\xi(t)=t$ gives
a trace that moves outward to infinity, but stays within fixed distance from
the real axis. We also solve explicitly a more general version of the
evolution equation, in which~$\xi(t)$ is a superposition of the values~$\pm1$.


\end{abstract}

\section{Introduction}
\label{Introduction}

The Loewner differential equation was introduced by Karl L\"owner~\cite{LOE}
(who later changed his name to Charles Loewner) to study properties of
univalent functions on the unit disk. The differential equation is driven
by a function that encodes in an ingeneous way continuous curves slitting
the disk from the boundary. A few years ago, Oded Schramm realized that
by taking Brownian motion as the driving function, the Loewner equation
generates families of random curves with a conformally invariant measure,
that he called ``Stochastic Loewner Evolutions'' (SLE's)~\cite{SCH}. Moreover,
he showed that if the loop-erased random walk has a scaling limit, and if
this scaling limit is conformally invariant, then it must be described by
an SLE, and he made similar conjectures about the scaling limits of uniform
spanning trees and of critical percolation. The existence of conformally
invariant scaling limits and the connections with SLE were established for
the first two models by Lawler, Schramm and Werner~\cite{LSW2}, and for
critical site percolation on the triangular lattice by Smirnov~\cite{SMI}.
Over the past few years Schramm's idea has been investigated further, mainly
in a combined effort of Lawler, Schramm and Werner~\cite{LSW}, leading to
new developments in the study of (the scaling limit of) discrete models.

In SLE an ensemble of driving functions generates an ensemble of shapes. It is
a difficult task to understand fully the relation between these two ensembles.
For this reason we investigate here the simpler problem of the relation
between a driving term in Loewner's equation and the curve it generates, by
explicit calculation. In particular we study exact, deterministic solutions
of Loewner's equation in the upper half~$\mathcal{H}$ of the complex plane.
One of the purposes of this study is to present some explicit examples that
may help to elucidate the behavior of the curves generated by Loewner's
equation. For example, our solution for square-root forcing in
section~\ref{SquareRootForcing} shows a transition from a simple curve that
stays above the real line to a curve intersecting the real line, which is
similar to the transition occurring at $\kappa=4$ in SLE.

\begin{figure}
 \centering\includegraphics{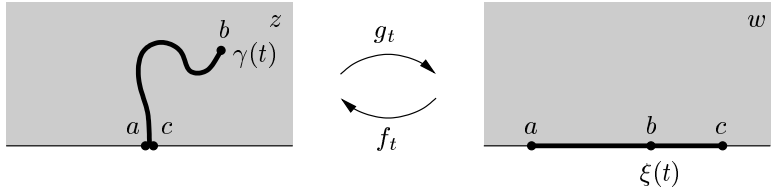}
 \caption{This figure shows how a slit~$\gamma$ in the $z$-plane maps into
  a linear boundary in the $w$-plane. Notice that two neighbouring points
  in the $z$-plane have images which are far apart in the $w$-plane.}
 \label{slitmap}
\end{figure}

To get us started, we need to understand how the Loewner equation describes
curves slitting the half-plane~$\mathcal{H}$. So suppose that $\gamma(t)$,
$t\geq0$, is a simple curve in $\mathcal{H}$ emanating from the real line.
Then according to conformal mapping theory, there exists a unique conformal
mapping of the form $w=g_t(z)$ that takes the domain~$\mathcal{R}$ consisting
of all points in~$\mathcal{H}$ minus those on the slit~$\gamma[0,t]$ onto the
upper half of the $w$-plane, in such a way that near infinity this map has
the expansion
\begin{equation}
 g_t(z) = z + c(t)/z + O(1/z^2).
 \label{EXPANSION}
\end{equation}
See figure~\ref{slitmap} for a schematic picture of the mapping. It is a
theorem that the coefficient $c(t)$ (which is called the \emph{capacity})
is continuously increasing with~$t$. Therefore, the parameterization of
the curve can be chosen such that $c(t)=2t$, and henceforth we assume that
this is the case.

The conformal maps $g_t$ satisfy a surprisingly simple differential equation,
which is the Loewner equation
\begin{subequations}
\label{LE}
\begin{equation}
 \frac{dg_t}{dt}=\frac{2}{g_t-\xi(t)}
\end{equation}
with the initial condition
\begin{equation}
 g_0=z
\end{equation}
\end{subequations}
for all $z$ in $\mathcal{H}$. Here, the value of $\xi$ at time $t$ is just
the image of the point $\gamma(t)$ under the map~$g_t$. We call $\xi(t)$ the
driving function or forcing function of the Loewner equation. It is continuous
and real-valued. (For a text-book discussion of Loewner's equation and proofs
of the above statements when the domain is the unit disk, see \cite{AHL2}
and~\cite{DUR}. The half-plane case we are considering here is analogous and
is discussed in~\cite{LAW} and~\cite{LSW1}).

Conversely, the Loewner equation~(\ref{LE}) has a solution for any given
continuous real-valued function~$\xi(t)$. This generates the function~$g_t(z)$.
For each value of $t$ this function can be thought of as a mapping of the
form $w=g_t(z)$ which takes some connected subset of the upper-half $z$-plane,
$\mathcal{R}$, into the entire region~$\mathcal{H}$ above the real axis of
the $w$-plane. Correspondingly, there exists an inverse function~$f_t(w)$,
which obeys $g_t(f_t(w))=w$ for all~$w$ in the upper half-plane. This function
maps the region~$\mathcal{H}$ of the $w$-plane into the region~$\mathcal{R}$
of the $z$-plane. It obeys the partial differential equation
\begin{equation}
 \frac{\partial f_t(w)}{\partial t} = -\frac{2}{w-\xi(t)}
 \frac{\partial f_t(w)}{\partial w}.
 \label{LEI}
\end{equation}

The Loewner equation continually generates new singular points of~$g_t$
that are mapped by~$g_t$ onto the corresponding points~$\xi(t)$ in the
$w$-plane. Thus, as time goes on, points are continually removed
from~$\mathcal{R}$. If~$\xi(t)$ is sufficiently smooth~\cite{MR}, these
singular points form a simple curve~$z_c(t)$, where the points~$z_c(t)$ obey
\begin{equation}
 g_t(z_c(t))=\xi(t).
 \label{SING}
\end{equation}
This curve is called the trace of the Loewner evolution, but we also refer
to it as the line of singularities. (Naturally, if we take~$\xi(t)$ to be
the driving function corresponding to a given curve~$\gamma(t)$ as described
above, then $z_c(t)$ will coincide with~$\gamma(t)$).

Equations~(\ref{LE}) have a simple scale invariance property under changing
both the scale of space and of time. Any change of time of the form
$t\mapsto\alpha^2t$ can be compensated by a change of scale. To do this,
define a new~$g$ and~$\xi$ by
\begin{equation}
 \tilde{g}_t(z) := \frac{1}{\alpha}g_{\alpha^2 t}(\alpha z)
 \quad\mbox{and}\quad
 \tilde{\xi}(t) =  \frac{1}{\alpha}\xi(\alpha^2 t).
 \label{SCALE}
\end{equation}
This pair then satisfies equations~(\ref{LE}). Note that by the Schwarz
reflection principle of complex analysis~\cite{AHL1}, the map~$g_t$ extends
to the lower half-plane and satisfies $g_t(\bar{z})=\overline{g_t(z)}$.
It follows from equation~(\ref{SCALE}) that flipping the sign of the forcing
has the effect of reflecting the regions $\mathcal{R}$ and~$\mathcal{H}$
(and hence also the line of singularities~$z_c(t)$) in the imaginary axis.
Likewise, a shift in the driving function~$\xi(t)$ can be compensated by
a similar shift in~$g_t$, since the pair
\begin{equation}
 \tilde{g}_t(z) := g_{t}(z-\alpha)+\alpha
 \quad\mbox{and}\quad
 \tilde{\xi}(t) =  \xi(t)+\alpha
 \label{SHIFT}
\end{equation}
again solves equations~(\ref{LE}). This shows that a shift in the driving
function simply produces a shift of the same magnitude in the trace~$z_c(t)$.

In this note we are interested in the question how the behavior of the trace
is related to that of the driving function. This question can be investigated
both numerically and, in some cases, by exact methods. A simple numerical
method  to find the trace generated by a given driving function~$\xi(t)$, is
to numerically integrate Loewner's equation backwards from the initial
condition $g_t=\xi(t)$ to obtain the value of $g_0=z_c(t)$. To deal with the
singularity in Loewner's equation, we can use the first-order approximation
\begin{equation}
 g_t = g_{t-dt} + \frac{2\,dt}{g_{t-dt}-\xi(t-dt)}.
\end{equation}
Substituting $g_t=\xi(t)$ and solving this equation for $g_{t-dt}$ gives
an approximation for the value of~$g$ at time~$t-dt$, from where we then
integrate backwards to obtain~$g_0$. This method is good enough for the
purpose of producing pictures. The figures \ref{linear}, \ref{spirals},
\ref{intersections} and~\ref{doubletrace} in this note were obtained in
this way.

Exact solutions of equations~(\ref{LE}) can be  found in a few cases. The
simplest are ones in which the forcing has one of the forms
\begin{subequations}
 \begin{equation}
  \xi(t) = C (1-t)^\beta
 \label{FORCEa}
 \end{equation}
or
 \begin{equation}
  \xi(t) = C t^\beta
 \label{FORCEb}
 \end{equation}
\end{subequations}
for $\beta=0$, $1/2$, and $1$. For a driving function of the form of
equation~(\ref{FORCEb}), a rescaling according to~(\ref{SCALE}) gives the
new forcing 
\begin{equation}
 \tilde{\xi}(t) = C \alpha^{2\beta-1} t^\beta.
\end{equation}
This result allows us to scale away the constant~$C$ when $\beta\neq1/2$. For
the special value $\beta=1/2$ the multiplicative constant can not be scaled
away, and can be important in determining the form of the solution. 

In sections~\ref{ConstantForcing}, \ref{LinearForcing}
and~\ref{SquareRootForcing} we shall derive the solutions respectively for
the powers $0$, $1$, and~$1/2$. Section~\ref{HalfCircle} contains an
explicit construction of the maps~$g_t$ for a half-plane slit by an
arc, and we point out that this is a special case of square-root forcing with
a finite-time singularity. In section~\ref{VaryingForcing} we consider a more
general version of Loewner's equation, where the forcing is a superposition
of the values~$\pm1$ and produces two lines of singularities.

\section{Constant Forcing}
\label{ConstantForcing}

Here we look at the almost trivial situation in which $\beta=0$ and we can
take
\begin{equation}
 \xi(t)=A.
\end{equation}
This situation is well-known (see for example~\cite{LAW} and appendix~A
in~\cite{BAU}), but for the sake of completeness we also treat it here. We
could of course shift~$A$ away by using equation~(\ref{SHIFT}), but this
case is so easy that shifting is hardly worth the effort. The equation
for~$g$ can be solved by inspection, giving
\begin{equation}
 g_t(z)=A +[(z-A)^2+4t]^{1/2}.
 \label{gC}
\end{equation} 
The inverse map is of the same form as the direct one, namely 
\begin{equation}
 f_t(w)=A +[(w-A)^2-4t]^{1/2}.
\end{equation}

\begin{figure}
 \centering\includegraphics{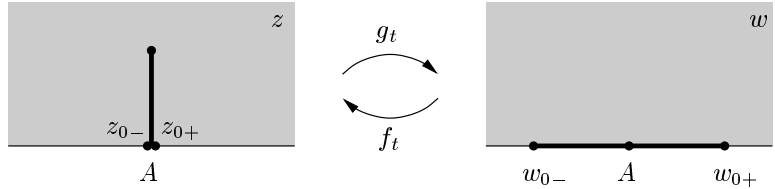}
 \caption{The trace produced by a constant forcing is a vertical line
  segment.}
 \label{constant}
\end{figure}

At time~$t$, the function~$g$ acquires a new singularity at the point
\begin{equation}
 z_c(t)=A+2it^{1/2}
\end{equation}
which maps under~$g_t$ to the point~$\xi(t)=A$ in the $w$-plane. Thus the
singularities form a line segment parallel to the imaginary axis and the
mapped region of the $z$-plane is the upper half-plane minus that line
segment. The mapping has a singular point at the place where the trace meets
the real axis. We can approach this point $z_0=A$ from two different
directions, giving us two different images at the points 
\begin{equation}
 w_{0\pm} = A \pm 2t^{1/2}.
\end{equation}
Figure~\ref{constant} gives an illustration of the mapping.

Because any smooth enough function~$\xi(t)$ looks locally as if it were
constant, we might expect that smooth~$\xi(t)$'s generate traces of
singularities which look like simple curves coming up from the real axis.
In fact, one can prove that this statement is correct whenever~$\xi(t)$
is sufficiently smooth so that it is H\"older continuous with exponent~$1/2$
and sufficiently small norm, i.e.\ so that
\begin{equation}
 \lim_{s\downarrow0} \frac{|\xi(t-s)-\xi(t)|}{s^{1/2}} < C
 \label{COND}
\end{equation}
for all~$t$ and some constant~$C$. The tricky issue is, when can it be that
the curve will touch itself or hit the real line. Condition~(\ref{COND}) is
enough to ensure that this will not happen, as was shown by Marshall and
Rohde~\cite{MR}, who did not give the value of the constant~$C$. Our solution
in section~\ref{SquareRootForcing} below gives a specific example of the
transition occuring at $C=4$. In a parallel and independent paper,
Lind~\cite{LIN} proves also generally that H\"older-$1/2$ functions with
coefficients $C<4$ generate slits.

\section{Linear Forcing}
\label{LinearForcing}

The next case, not much harder, has the forcing increase linearly with
time according to
\begin{equation}
 \xi(t)=t.
 \label{Linear}
\end{equation}
We do not need an additive or multiplicative constant in
equation~(\ref{Linear}) because these can be eliminated by equations
(\ref{SCALE}) and~(\ref{SHIFT}). If we now redefine the independent
variable to be $h=g-\xi$, then~$h$ obeys
\begin{equation}
 \frac{dh}{dt}=\frac{2-h}{h}=:-\frac{dh}{dF(h)}
\end{equation}
where~$F(h)$ is the function
\begin{equation}
 F(h)=h+2 \ln(2-h).
\end{equation}
In terms of $F$ the solution is
\begin{equation}
 F(h)=-t +c(z).
\end{equation}
Here, $c(z)$  is a constant which must be set from the initial condition,
$g_0=z$, giving the solution $F(h)=-t+F(z)$ or equivalently
\begin{equation}
 F(g-t)=F(z)-t.
 \label{Csoln}
\end{equation}

To find the trajectory of the singularity, use equations (\ref{SING})
and (\ref{Csoln}) to get
\begin{equation}
 F(z_c(t))=F(0)+t=2\ln2 +t.
 \label{Csoln2}
\end{equation}
It is possible to find an explicit expression for this trajectory. To do so,
substitute $2-z_c(t)=r_t\exp(-i\phi_t)$ into the previous equation. Then
split the equation in its real and imaginary parts to obtain
\begin{subequations}
\begin{eqnarray}
 && 2\ln r_t - r_t\cos\phi_t = 2\ln2 + t - 2, \\
 && r_t = 2\phi_t / \sin\phi_t.
\end{eqnarray}
\end{subequations}
After substituting the second equation into the first, it can be shown
that~$\phi$ increases monotonously in time from the value $\phi_0=0$ to
$\phi_\infty=\pi$.

\begin{figure}
\centering\includegraphics{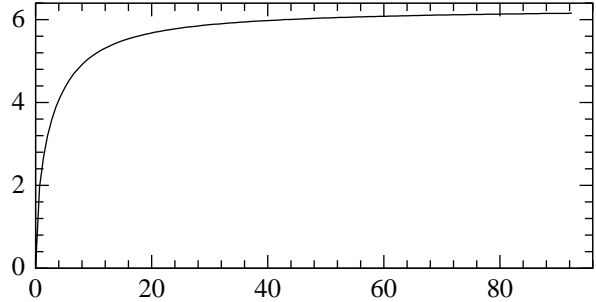}
\caption{The shape of the line of singularities in the $z$-plane that arises
from the forcing $\xi(t)=t$, up to time $t=100$.}
\label{linear}
\end{figure}

In terms of the parameter~$\phi$, the line of singularities may be written
explicitly as
\begin{equation}
 z_c(t) = 2 - 2\phi_t\cot\phi_t + 2i\phi_t.
\end{equation}
This shows that the line of singularities moves outward to infinity while
remaining within a fixed  distance of the real axis. Figure~\ref{linear}
shows the first part of this trajectory.

For small and large values of~$t$ we can also construct asymptotic forms
of the solution~(\ref{Csoln2}). As~$t$ goes to zero, one can expand
$F(z_c(t))$ to third order in $z_c(t)$ and find that, to this order, the
result is
\begin{equation}
 z_c(t)=2 i t^{1/2} + \frac{2}{3}t + O(t^{3/2}).
\end{equation}
For large~$t$,
\begin{equation}
 z_c(t) =t- 2\ln[(t-2)/2]+2\pi i + O(\ln t/t).
\end{equation}

\section{Square-root Forcing}
\label{SquareRootForcing}

The next case, much more interesting, has the forcing be a square-root
function of time. This case has two subcases: one in which the forcing has
a finite-time singularity based on the rule of equation~(\ref{FORCEa}), i.e.
\begin{equation}
 \xi(t) = 2 [\kappa (1- t)]^{1/2}, \qquad t\leq1, \qquad \kappa\geq 0.
 \label{FORCEap}
\end{equation}
Here, as we shall see, the result changes qualitatively as~$\kappa$ varies,
with the critical case being $\kappa=4$ where the singular curve intersects
the real axis. In the other situation, the forcing has an infinite-time
singularity with the rule
\begin{equation}
 \xi(t) = 2[\kappa t]^{1/2}, \qquad \kappa\geq 0.
 \label{FORCEbp}
\end{equation}
In this case $z_c(t)$ is just a straight line, as is already clear from the
scaling relation~(\ref{SCALE}).

\subsection{Infinite-time Singularity}

To find the solution for the forcing~(\ref{FORCEbp}), define the new variable
$G=g/t^{1/2}$ and set $\tau=\ln t$. Then~$G$ satisfies
\begin{equation}
 \frac{d G}{d\tau} = -G/2 + 2/(G-2\kappa^{1/2})
  = \frac{(G-y_+)(G-y_-)}{2(2\kappa^{1/2}-G)}
\end{equation}
where $y_{\pm} = \kappa^{1/2}\pm(\kappa+4)^{1/2}$. It follows that
\begin{equation}
 \frac{d G}{d\tau}\left[ \frac{y_-}{G-y_+}-\frac{y_+}{G-y_-} \right]
  = \frac{1}{2}(y_+ - y_-).
\end{equation}
Therefore, if we set
\begin{equation}
 H(G):= \frac{2y_+\ln(G-y_-) - 2y_-\ln(G-y_+)}{y_+ - y_-}
\end{equation}
then $dH(G)/d\tau=-1$ which integrates to
\begin{equation}
 -H(G) = \tau + c(z) = \ln t + c(z).
 \label{FORCEbpSoln}
\end{equation}

The constant~$c(z)$ can be determined from the observation that in the
limit when~$t$ approaches~$0$,
\begin{eqnarray}
 H(G) + \ln t &=&
  \frac{2y_+\ln(g-{y_-}t^{1/2}) - 2y_-\ln(g-{y_+}t^{1/2})}{y_+ - y_-}
  \nonumber\\
  &\longrightarrow& 2\ln z.
\end{eqnarray}
Our solution~(\ref{FORCEbpSoln}) for general~$t$ becomes simply
\begin{equation}
 H(g/t^{1/2}) = 2\ln(z/t^{1/2}).
\end{equation}
Since the  line of singularities is determined by the condition that~$g$ is
equal to the forcing we get
\begin{equation}
 z_c(t) =B\,t^{1/2} \quad\mbox{where}\quad B=\exp[\frac{1}{2}H(2\kappa^{1/2})].
\end{equation}
More explicitly the coefficient is
\begin{equation}
 B = 2
  \left(\frac{(\kappa+4)^{1/2}+\kappa^{1/2}}
  				{(\kappa+4)^{1/2}-\kappa^{1/2}}
  \right)^{\displaystyle\frac{1}{2}\frac{\kappa^{1/2}}{(\kappa+4)^{1/2}}}
  \exp\left[\frac{1}{2}\pi i
            \left(1-\frac{\kappa^{1/2}}{(\kappa+4)^{1/2}}\right)
      \right],
\end{equation}
so that the line of singularities is set at an angle to the real axis which
is
\begin{equation}
\theta=\frac{1}{2}\pi
 \left(1-\frac{\kappa^{1/2}}{(\kappa+4)^{1/2}}\right).
\end{equation}
For $\kappa=0$ the line is (as we know) perpendicular to the real axis while
as $\kappa\rightarrow\infty$ the angle of intersection becomes smaller and
smaller.

\subsection{Finite-time Singularity}

Now we turn to the forcing~(\ref{FORCEap}) with a singularity after a finite
time. In this case, one can eliminate the time-dependence by using the new
variable
\begin{equation}
 G= g/(1- t)^{1/2}.
\end{equation}
Then~$G$ obeys
\begin{equation}
 \frac{d G}{d \tau} = G/2+2/(G-2\kappa^{1/2})
\end{equation}
where $\tau = - \ln (1-t)$. Once again the derivative becomes a simple
function of the unknown, again appearing in the form of a ratio of
polynomials, i.e.
\begin{equation}
 -\frac{d G}{d \tau} =\frac{(G-y_+)(G-y_-)}{2(2\kappa^{1/2}-G)}.
 \label{Geqn}
\end{equation}
where this time the roots are $y_\pm=\kappa^{1/2}\pm(\kappa-4)^{1/2}$. Notice
that now the roots are real and positive when $\kappa\ge 4$ but that they are
complex for $\kappa<4$. Thus we can expect a qualitative change in the
solution at $\kappa=4$.

The integration of the equation proceeds exactly as in the previous case,
giving the solution
\begin{equation}
 H(G)=\tau+c(z)
\end{equation}
where~$H$ has the same form as before, but with different~$y_\pm$:
\begin{equation}
 H(G)= \frac{2y_+\ln(G-y_-) - 2y_-\ln(G-y_+)}{y_+ - y_-}.
\end{equation}
The initial condition then implies that
\begin{equation}
 H(g/(1- t)^{1/2})=-\ln(1-t)+H(z)
\end{equation}
and the equation for the line of singularities is then
\begin{equation}
 H(2 \kappa^{1/2})=-\ln(1-t)+H(z_c(t)).
 \label{SQRTsoln}
\end{equation}
Note that equation~(\ref{SQRTsoln}) for the line of singularities tells us
that $H(z_c(t))$ must approach~$-\infty$ when~$t$ approaches~$1$. In the
following subsections we shall consider the asymptotics of~$z_c(t)$ in
this limit in more detail. We separate the discussion into the two cases
$\kappa<4$ and $\kappa>4$. Then the section is completed with the derivation
of the shape of the critical line of singularities at $\kappa=4$.

\subsubsection[The logarithmic Spiral for kappa < 4]{The logarithmic Spiral for $\kappa<4$}

\begin{figure}
 \centering\includegraphics{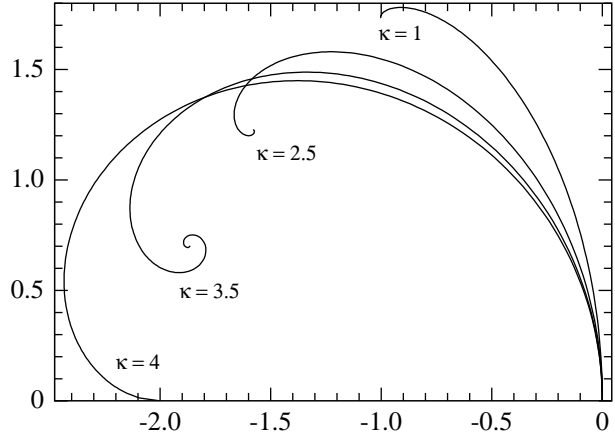}
 \caption{The trace of singularities for square-root forcing with different
  values of $\kappa<4$. Also shown is the limiting case $\kappa=4$ where the
  trace starts to hit the real line. All the curves have been translated to
  make them start in the origin at time~$t=0$.}
 \label{spirals}
\end{figure}

When $\kappa<4$, the real and imaginary parts of~$H(z)$ are
\begin{subequations}
\begin{eqnarray}
 \Re H(z) &=&
  \frac{\kappa^{1/2}}{(4-\kappa)^{1/2}}\Arg(z-y_-) + \ln|z-y_-|
   \nonumber\\ && \null
  -\frac{\kappa^{1/2}}{(4-\kappa)^{1/2}}\Arg(z-y_+) + \ln|z-y_+|,\\
 \Im H(z) &=& \Arg(z-y_-) - \frac{\kappa^{1/2}}{(4-\kappa)^{1/2}}\ln|z-y_-|
  \nonumber\\ && \null
  +\Arg(z-y_+) + \frac{\kappa^{1/2}}{(4-\kappa)^{1/2}}\ln|z-y_+|,
\end{eqnarray}
\end{subequations}
from which it follows easily that $H(2\kappa^{1/2})$ is a positive real
constant. Since the real part of $H(z_c(t))$ must go to~$-\infty$ when~$t$
approaches~$1$, we conclude that $z_c(1)=y_+$, since the other possible
candidate for the limit, i.e.\ $y_-$, is in the wrong half-plane. It then
follows from the observation that $\Im H(z_c(t))$ must vanish, that the
trace must have wrapped around~$y_+$ an infinite number of times. Hence
the line of singularities is spiralling in towards the point~$y_+$, as is
depicted in figure~\ref{spirals}.

To find a more explicit expression for the asymptotics, notice that
when $z_c(t)$ is close to~$y_+$, we can make the approximations
\begin{equation}
 \ln|z_c(t)-y_-| \approx \ln\left(2(4-\kappa)^{1/2}\right) \mbox{\ and\ }
 \Arg(z_c(t)-y_-) \approx \pi/2.
\end{equation}
Splitting equation~(\ref{SQRTsoln}) in its real and imaginary parts, and
substituting the above approximate values then gives a system of two
equations that we can solve for the unknows $\ln|z_c(t)-y_+|$ and
$\Arg(z_c(t)-y_+)$. This gives the result
\begin{subequations}
\begin{eqnarray}
 \ln|z_c(t)-y_+|
  &\approx& \big[ A(\kappa) + (4-\kappa)\ln(1-t) \big] / 4, \\
 \Arg(z_c(t)-y_+)
  &\approx& \big[ B(\kappa) - \kappa^{1/2}(4-\kappa)^{1/2}\ln(1-t) \big] / 4,
\end{eqnarray}
where the constants are
\begin{eqnarray}
 A(\kappa) &=& \ln16+(\kappa-2)\ln(4-\kappa) \nonumber\\ && \null
  + (2\Arg(y_+)-\pi)\kappa^{1/2}(4-\kappa)^{1/2}, \\
 B(\kappa) &=& \kappa^{1/2}(4-\kappa)^{1/2}\ln(4-\kappa)
  \nonumber\\ && \null
  + (\pi-2\Arg(y_+))\kappa - 2\pi.
\end{eqnarray}
\end{subequations}
Thus, the distance between $z_c(t)$ and~$y_+$ is decreasing like a power
law in~$t$, whereas the winding number of $z_c(t)$ around~$y_+$ grows only
logarithmically.

\subsubsection[The Intersection for kappa > 4]{The Intersection for $\kappa>4$}

\begin{figure}
 \centering\includegraphics{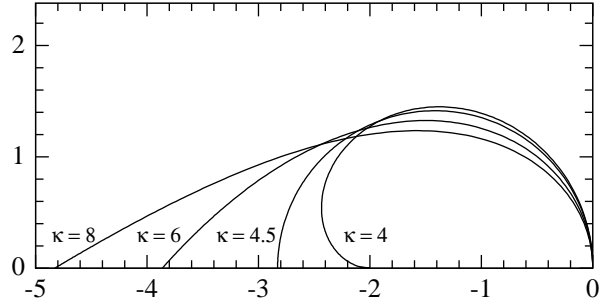}
 \caption{For a square-root forcing with $\kappa\geq4$ the line of
  singularities intersects the real line at an angle that varies with
  $\kappa$. The curves were translated to make them start in the origin.}
 \label{intersections}
\end{figure}

For $\kappa>4$ the real and imaginary parts of~$H(z)$ are
\begin{subequations}
\begin{eqnarray}
 \Re H(z) &=&
 \left(1+\frac{\kappa^{1/2}}{(\kappa-4)^{1/2}}\right)\ln|z-y_-|
 \nonumber\\ && \null
 + \left(1-\frac{\kappa^{1/2}}{(\kappa-4)^{1/2}}\right)\ln|z-y_+|, \\
 \Im H(z) &=&
 \left(1+\frac{\kappa^{1/2}}{(\kappa-4)^{1/2}}\right)\Arg(z-y_-)
 \nonumber\\ && \null
 + \left(1-\frac{\kappa^{1/2}}{(\kappa-4)^{1/2}}\right)\Arg(z-y_+),
\end{eqnarray}
\end{subequations}
and $H(2\kappa^{1/2})$ is again a positive real number. This time, as~$t$
approaches one, $z_c(t)$ must approach the point~$y_-$ on the real line,
see figure~\ref{intersections}. In fact, we can calculate the angle at
which the line of singularities intersects the real line from the
observation that $\Im H(z_c(t))$ must vanish.

Indeed, if we denote by~$\phi$ the angle at which~$z_c(t)$ hits the real
line (measured from the real line in the positive direction), then we have
that in the limit $t\rightarrow1$, $\Arg(z_c(t)-y_+)\rightarrow\pi$ whereas
$\Arg(z_c(t)-y_-)\rightarrow\phi$. The condition $\Im H(z_c(t))=0$ then
gives
\begin{equation}
 \phi =
 \pi\frac{\kappa^{1/2}-(\kappa-4)^{1/2}}{\kappa^{1/2}+(\kappa-4)^{1/2}}
\end{equation}
when~$t$ goes to~$1$. As we can see, we have a glancing incidence for
$\kappa=4$ and $\kappa\rightarrow\infty$, while for $\kappa=9/2$ the
incidence is perpendicular. In fact we can prove that for $\kappa=9/2$
the line of singularities is a half-circle of radius~$\sqrt{2}$. We will do
this in section~\ref{HalfCircle}, where we give an explicit construction
of the conformal maps~$g_t$ when the half-plane is slit by a half-circle.

\subsubsection[The critical Curve at kappa = 4]{The critical Curve at $\kappa=4$}

It is not difficult to find an explicit expression for the critical line
of singularities at $\kappa=4$. In this case, the two roots in
equation~(\ref{Geqn}) are the same, $y_\pm=2$, and the equation can be
rewritten as
\begin{equation}
 \left[\frac{2}{G-2}-\frac{4}{(G-2)^2}\right]\frac{dG}{d\tau}=1.
\end{equation}
Integration is straightforward and gives
\begin{equation}
 2\ln(G-2)+\frac{4}{G-2}=-\ln(1-t)+2\ln(z-2)+\frac{4}{z-2}
\end{equation}
where we have determined the integration constant from the initial condition
$G_0=g_0=z$. The line of singularities is determined by the condition that
for $z=z_c(t)$, $G=4$.

Substituting $z_c(t)-2=r_t\exp(i\phi_t)$ and splitting the equation in real
and imaginary parts leads to the expression
\begin{equation}
 z_c(t)=2+\frac{\sin2\phi_t}{\phi_t}+2i\frac{\sin^2\phi_t}{\phi_t}
\end{equation}
for the critical curve in terms of the parameter~$\phi$. This parameter~$\phi$
increases monotonously with time from $\phi_0=0$ to $\phi_1=\pi$.

\section{Loewner Evolution for a growing Arc}
\label{HalfCircle}

\begin{figure}
 \centering\includegraphics{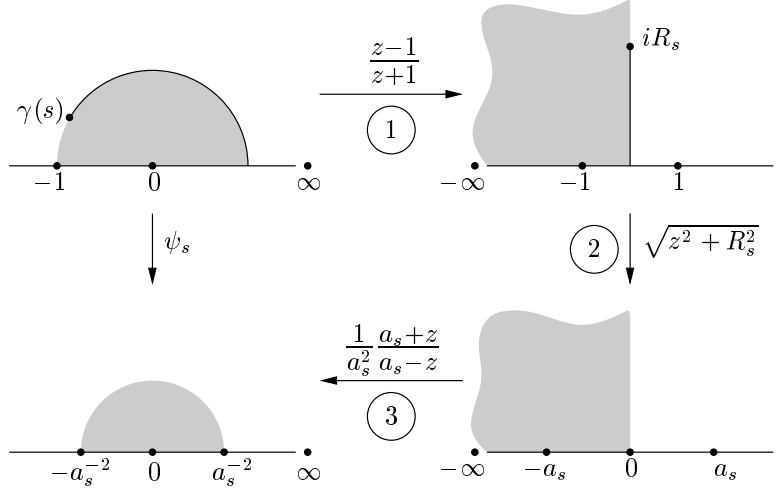}
 \caption{The construction of the map~$\psi_s$ taking~$\mathcal{H}$ minus
  the slit $\gamma[0,s]$ onto~$\mathcal{H}$, in three steps. The illustration
  also tracks the images of the upper half of the unit disk and of the four
  ``special'' points $-1$, $0$, $\gamma(s)$ and~$\infty$. The definitions
  of $\gamma(s)$, $R_s$ and~$a_s$ are in the text.}
 \label{halfcircle}
\end{figure}

In this section, we consider the situation where the upper
half-plane~$\mathcal{H}$ is slit by the curve $\gamma(s)=\exp(is)$,
$s\in\left[0,\pi\right)$. The idea is to construct the conformal maps that
map the half-plane minus the slit $\gamma[0,s]$ back to the half-plane and
have the expansion of equation~(\ref{EXPANSION}). After reparameterizing time
one can then identify the driving function, and verify that the obtained
maps~$g_t$ satisfy Loewner's equation. The form of the normalized maps and
of the driving function already appears in appendix~A of~\cite{BAU}, but for
completeness we present a short account of the derivation of these results
below. A simple rescaling argument then allows us to make the connection with
section~\ref{SquareRootForcing} and prove that the square-root forcing we
considered there generates a half-circle of radius~$\sqrt{2}$ when
$\kappa=9/2$.

To obtain the normalized conformal maps, we start by constructing the
map~$\psi_s$ of figure~\ref{halfcircle} as the composition of the three maps
depicted there. The explicit form of this map is
\begin{equation}
 \label{PSI}
 \psi_s(z) = \frac{1}{a_s^2}\frac{a_s+f_s(z)}{a_s-f_s(z)}
  \qquad\mbox{for\ }z\in\mathcal{H}\setminus\gamma[0,s]
\end{equation}
where $a_s:=\cos(s/2)^{-1}$ and the function~$f_s(z)$ is given by
\begin{equation}
 \label{F}
 f_s(z) = \sqrt{\Big(\frac{z-1}{z+1}\Big)^2+R_s^2}
        = \sqrt{a_s^2-\frac{4z}{\smash{(z+1)^2}}},
\end{equation}
with $R_s:=\tan(s/2)$. We can now expand~$\psi_s$ for large~$z$, and
identify the capacity~$c(s)$. A simple translation will then give us the
map satisfying~(\ref{EXPANSION}).

Expanding~$\psi_s(z)$ for large~$z$ up to order~$1/z$ gives
\begin{equation}
 \psi_s(z) = z + (2-2a_s^{-2}) + (1-a_s^{-4})/z + O(1/z^2).
\end{equation}
Therefore, the map $\psi_s+2a_s^{-2}-2$ has the expansion of
equation~(\ref{EXPANSION}), with capacity $c(s)=1-a_s^{-4}$. The proper time
reparameterization that makes the capacity equal to~$2t$ is given by
\begin{equation}
 2t(s) := 1-a_s^{-4} = 1-\cos^4(s/2)
 \quad\mbox{for}\quad s\in\left[0,\pi\right)
\end{equation}
which has the inverse
\begin{equation}
 s(t)/2 = \arccos\left((1-2t)^{1/4}\right)
 \quad\mbox{for}\quad t\in\left[0,1/2\right).
\end{equation}
The maps~$g_t$ are then given by $g_t(z)=\psi_{s(t)}(z)+2a_{s(t)}^{-2}-2$.
More explicitly, this can be written in the form
\begin{equation}
 g_t(z) = \frac{2(z-1)^2+4z(1-2t)^{1/2}+
  2(z+1)\sqrt{(z+1)^2-4z(1-2t)^{1/2}}}{4z}.
 \label{CIRCLEMAP}
\end{equation}
We know from figure~\ref{halfcircle} that this map takes the point
$\gamma(s(t))$ onto the image $3a_{s(t)}^{-2}-2$, which implies that the
driving function must be
\begin{equation}
 \xi(t) = 3(1-2t)^{1/2}-2.
\end{equation}
Indeed, one can check by differentiating~(\ref{CIRCLEMAP}) with respect to
time, that~$g_t$ satisfies Loewner's equation with this driving function.

To make the connection with section~\ref{SquareRootForcing}, we note that
a rescaling of the form~(\ref{SCALE}) with $\alpha=1/\sqrt{2}$ turns the
driving function into a function of the form of equation~(\ref{FORCEap})
with $\kappa=9/2$, up to a translation. We conclude that this particular
square-root forcing produces a line of singularities that is a half-circle
of radius~$\sqrt{2}$.

We conclude this section with a discussion of its relation to an old paper
by Kufarev~\cite{KUF}. Kufarev considers the equivalent of Loewner's
equation~(\ref{LE}) in the unit disk, and gives an explicit solution of this
equation for negative times. A half-plane version of Kufarev's example is
obtained as follows. Suppose that we set $\xi(-s)=3\sqrt{2s}$ for~$s\geq0$.
Then it can be verified that the solution of~(\ref{LE}) at time~$-s$ is the
normalized map that takes~$\mathcal{H}$ onto $\mathcal{H}$ minus the half-disk
of radius~$\sqrt{2s}$ centered at~$2\sqrt{2s}$. Observe that for~$s=1/2$, up
to a translation, this solution is just the inverse of~(\ref{CIRCLEMAP}) at
time~$t=1/2$. This is not a coincidence because generally, if one sets
$\xi(-t):=\xi(T-t)$ and solves Loewner's equation for both $g_T$ and~$g_{-T}$,
the two results are related by $g_{-T}=g_T^{-1}$.

\section{Multivalued Forcing}
\label{VaryingForcing}

In this section we look at a special case of the more general version of
Loewner's equation (appearing for example in~\cite{LAW})
\begin{equation}
 \frac{dg_t}{dt} = \int\frac{2}{g_t-x}\,d\mu_t(x)
\end{equation}
where~$\mu_t$ is a measure on the real line that can be time-dependent. Here
we will take~$\mu_t$ to be time-independent and assigning a mass~$p_j$ to the
values~$\xi_j$ such that $\sum_j p_j=1$. Then the equation for~$g$ takes the
form
\begin{equation}
 \frac{dg}{dt}= \sum_j \frac{2 p_j}{g-\xi_j}.
 \label{JUMP}
\end{equation}
This can be seen as a case where the forcing~$\xi(t)$ is a superposition
of values, and we also believe that this can be described as a case where the
forcing~$\xi(t)$ makes rapid (random) jumps between the values~$\xi_j$.

Equations like~(\ref{JUMP}) are generally easily integrated. Take the specific
case in which the possible values of~$\xi$ are~$\pm1$, which are taken on with
equal probability. Then equation~(\ref{JUMP}) becomes
\begin{equation}
 \frac{dg}{dt}= \frac{2g}{g^2-1}
 \label{JUMP2}
\end{equation}
which then integrates to give
\begin{equation}
 g^2/2 -\ln g = 2t+z^2/2-\ln z.
 \label{JUMPsoln}
\end{equation}
There are two traces, starting respectively from $z=\pm1$ and then moving
upward towards~$i\infty$. The traces are found, as before, by setting~$g$
equal to the forcing. Thus the right-hand trace obeys
\begin{equation}
 z_{c+}(t)^2 - 2\ln z_{c+}(t) = 1-4t.
\end{equation}

\begin{figure}
 \centering\includegraphics{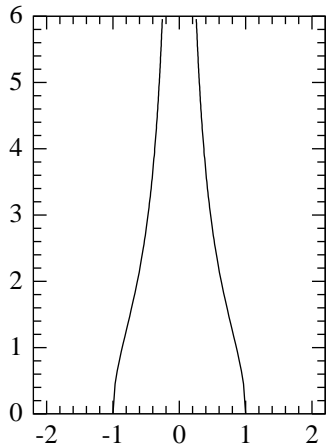}
 \caption{A forcing that is a superposition of the values $\pm1$ produces
  two lines of singularities that approach each other as time goes on. This
  illustration shows the two lines up to time $t=10$.}
 \label{doubletrace}
\end{figure}

Again we can obtain an explicit expression for the trace, like we did for
the case of linear forcing in section~\ref{LinearForcing}, and for the
critical curve in section~\ref{SquareRootForcing}. After substituting
$z_c(t)=r_t\exp(i\phi_t)$ and splitting the previous equation in real and
imaginary parts, we obtain
\begin{equation}
 z_{c\pm}(t) = \sqrt{\frac{2\phi_t}{\sin(2\phi_t)}}\,
  \Big(\pm\cos\phi_t+i\sin\phi_t\Big)
\end{equation}
where~$\phi$ increases with time from the value $\phi_0=0$ to
$\phi_\infty=\pi/2$. Note that we have filled in the results for both traces,
using the fact that they must be symmetrically placed about the imaginary
axis. We can also compute the asymptotics. For small values of~$t$ we obtain
\begin{equation}
 z_{c\pm}(t) = \pm1 + i\sqrt{2t} \mp\frac{1}{3}t + O(t^{3/2})
\end{equation}
while for large values of~$t$ the traces behave like
\begin{equation}
 z_{c\pm}(t) =i\sqrt{(4t-1)-\ln(4t-1)} \pm \frac{\pi}{2(4t-1)^{1/2}}
  + O(\ln t/t^{3/2}).
\end{equation}
We see that as the traces move upward, they also approach one another, as
is shown in figure~\ref{doubletrace}.

\subsubsection*{Acknowledgements}

One of us (LPK) would like to thank Leiden University and Professor Wim van
Saarloos for their hospitality for a part of the period in which the work was
performed. We have had useful conversations about this research with Paul
Wiegmann, Ilya Gruzberg, Isabelle Claus, Marko Kleine Berkenbusch, and John
Cardy. We thank Steffen Rohde and Joan Lind for sharing their recent work on
Loewner's equation with us. LPK would like to acknowledge partial support
from the US NSF's Division of Materials Research. WK is supported by the
Stichting FOM (Fundamenteel Onderzoek der Materie) in the Netherlands.

\end {document}